\documentclass[10pt,aps,superscriptaddress,twocolumn,prc,floatfix,showpacs,nofootinbib,longbibliography,showkeys]{revtex4-1}
\newif\ifdraft
\drafttrue 
\draftfalse
\usepackage{epsf,color,colordvi}
\usepackage{graphicx}
\usepackage{tikz}
\usepackage{amsmath,amssymb}
\usepackage{times}
\usepackage{comment}
\usepackage{placeins}
\usepackage[normalem]{ulem}
\usepackage{soul}
\usepackage{subfigure}
\usepackage{balance}

\newcommand{\ts}{\textstyle }
\newcommand{\ds}{\displaystyle }
\newcommand{\be}[1]{\begin{equation}\label{eq:#1}}
\newcommand{\ee}{\end{equation}}
\newcommand{\bs}{\begin{split} }
\newcommand{\es}{\end{split} }
\newcommand{\SE}{Schr\"odinger equation }

\newcommand{\etal}{\textit{et al.~}}

\begin{document}

\author{W. van Dijk}
\email[Email: ]{vandijk@physics.mcmaster.ca}
\affiliation{Physics Department, Redeemer University College, Ancaster, Ontario, Canada  L9K 1J4}
\affiliation{Department of Physics and Astronomy, McMaster University, Hamilton, Ontario. Canada L8S 4M1}
\author{F. M. Toyama}
\email[Email: ]{toyama@cc.kyoto-su.ac.jp}
\email{JSPS Research Fellow}
\affiliation{Department of Computer Science, Kyoto Sangyo University, Kyoto 603-8055, Japan}

\title{Decay of a quasi-stable quantum system and quantum backflow}

\begin{abstract} 

The decay of quasi-stable quantum system involves primarily an outgoing probability current density.  However, during the transition from exponential to inverse-power-law decay there are time intervals during which this current, although small, is inward.  In this paper this inward flow is associated with quantum backflow.  Furthermore substantial backflow exists for time-evolving free wave packets which are initially confined in space. \\

\ifdraft{\noindent Manuscript file: $~\sim$/2019/backflow/notes/paper\_v2.tex \\} \fi  
\noindent Date: \today

\end{abstract}

\maketitle

\section{Introduction}

In the early days after the discovery of radioactive decay it was noted that the predominant characteristic of the decay is the exponential decrease of the number of atoms in the source.  Elementary theory of decay would suggest an outgoing flux density that tracks the decay.  However, according to quantum theory the outgoing probability current density, although mainly exponentially decreasing, can display fluctuations in time, which cause it to be negative or inward over some (small) intervals of time.  In this paper we employ models of quantum quasi-stable systems to investigate such negative probability current densities, and their relation to the initial wave function and to the phenomenon of quantum backflow.  As quantum backflow is not dependent on the (presence of an) interaction, we discuss also a simpler model of the time evolution of a localized free particle in order to estimate upper limits to the amount of backflow.  In the study we use the exact solution to the time-dependent \SE developed by us~\cite{vandijk99,vandijk02} for the $\delta$-function interaction, and we derive the maximum backflow using the time-dependent solution of the free Schr\"odinger equation.

In 1961 Winter~\cite{winter61} discussed the evolution of a quasi-stationary quantum state in terms of a model involving the decay  of particle through a $\delta$-shell barrier.  This model constitutes an extremely simple simulation of a decaying quantum system such as an $\alpha$-emitting nucleus.  It allows a transparent analysis of exponential and nonexponential decay, whose features also take form in more realistic models.

The decay of a quasi-stable system is characterized by a nonescape or survival probability that behaves as a power law in time for very short times after the creation of the system, followed by a dominant exponential decay, and then by a long-time behaviour characterized as an inverse-power law of time, first pointed out by Khaflin~\cite{khaflin58}.   The $\delta$-shell model was further investigated by Garc\'ia-Calder{\'o}n and Peierls in 1976~\cite{garcia76}, who provided an analytic expression of the wave function inside the potential barrier.  More recently analytic wave functions both inside and outside the potential barrier were derived~\cite{cavalcanti99,vandijk99,vandijk02}.  The determination of the exact wave function was broadened to those due to potential barriers of different shapes and to potentials supporting bound states~\cite{vandijk02}.  The experimental observation of the nonexponential decay is more elusive.  It is pointed out that decays of radioactive isotopes are not suitable candidates since the energy released tends to be much larger than the width of the energy distribution and many exponential lifetimes need to have elapsed before nonexponential decay sets in~\cite{ramirez_jimenez19}.    Nevertheless, by measuring the luminescence decays of dissolved organic materials, Rothe \etal~\cite{rothe06} obtained the first experimental evidence for the long-time nonexponential decay.  Recently, Crespi~\etal using integrated photonics~\cite{crespi19} also observed the inverse-power law decay, as well as the quadratic short-time decay behaviour.

A common and striking feature of the analyses is that the nonescape and survival probabilities and the probability current density display, besides the characteristic behaviour in the three time intervals indicated above, variable behaviour in the regions of transition during which the power law changes to the exponential, and the exponential to the inverse-power law.   The existence of these fluctuations was first shown by Winter~\cite{winter61}, but only recently did they receive detailed scrutiny~\cite{ramirez_jimenez19}.  Fluctuations of this kind are the motivation of the study of this paper.

  Normally one expects the probability current density of the decaying system to flow outward.  In fact the decay of quasi-stable systems is often described in terms of Gamow functions, which are characterized as $e^{ikr}$ at the boundary of, and outside, the potential region.  These outward travelling waves have positive wave numbers.  However exact determination of the probability current density yields negative values during certain time intervals, indicating that although the system is decaying there are times when the probability of the particle being inside the barrier is increasing.  This counterintuitive notion is strikingly displayed using Bohm's quantum trajectories where one observes certain trajectories leaving and then reentering the potential region~\cite{nogami00}.  

A similarly counterintuitive notion exists in the phenomenon of quantum backflow.  This was first pointed out by Alcock in 1969~\cite{allcock69c} and analyzed in greater detail by Bracken and Melloy in 1994~\cite{bracken94}.  The early discussions involve wave packets consisting of components with positive wave numbers only, i.e., travelling in the positive $x$ direction, which however yield a negative probability current density over some time intervals at some spatial point, say the origin.  This means that the probability of the wave packet being to the left of the origin increases during these time intervals.  Bracken and Melloy showed that there is no limit on the size of the time interval during which the backflow occurs, but there is a limit on the increase of particle  probability on the left of the origin.  (See also Refs.~\cite{yearsley12,yearsley13}.)   In fact they suggest a quantum number independent of physical quantities such as mass, time, and Planck's constant, which limits the increase to less than 0.04.  This number, labelled $c_{mb}$, was subsequently refined to a more precise value~\cite{penz06}.

Recently it was shown that quantum backflow is a universal quantum effect; it does not only pertain to interaction-free systems, but can be extended to scattering involving short-range potentials~\cite{bostelmann17}.  Moreover, Goussev~\cite{goussev19} demonstrates the equivalence between quantum backflow of a wave packet consisting of nonnegative momentum components and the reentry problem in which a free wave packet initially confined to a semi-infinite line, but unconstrained in momentum space, evolves in time to lower and, for certain intervals, to raise the probability of being in the confined space region. 

In this paper we consider the appropriateness of identifying the negative probability current density of the decaying quantum system with quantum backflow.    
In Sec.~\ref{sec:02} we review the fluctuating probability current density of the $S$-wave quantum decaying system, especially the time intervals during which it is negative.      In Sec.~\ref{sec:03} we present a possible quantum backflow interpretation, followed by an analysis of the time evolution of a free wave packet in an $S$-wave partial state in Sec.~\ref{sec:04}.  We discuss the backflow of the free particle as an eigenvalue problem in Sec.~\ref{sec:05}, and present a summary and concluding comments in Sec.~\ref{sec:06}.

\section{Decay through a delta-function barrier}
\label{sec:02}
\setcounter{equation}{0}

Consider a particle of mass $m$ initially confined to a spatial region $r\in(0,a)$.  Beginning at time $t_0$ it is allowed to escape through a $\delta$-function barrier at $r=a$.   The wave function of such a particle is a solution of the 
 time-dependent Schr\"odinger equation  
\begin{equation}\label{eq:3.01}
i\dfrac{\partial\psi(r,\tau)}{\partial \tau} = \left[-\dfrac{\partial^2~}{\partial r^2} + V(r)\right]\psi(r,\tau), \ \ \ \psi(r,0)=\phi_n(r),
\end{equation}
where $\phi_n(r)$ is the initial wave function at $\tau=0$ and $0\le r\le a$.  For simplicity we choose  generic units of time so that $\tau=\hbar(t-t_0)/(2m)$, where $t_0$ is the (arbitrary) initial time and $m$ the mass of the emitted particle.    The potential barrier is $V(r) = (\lambda/a)\delta(r-a)$.   The wave equation applies to the $S$-wave partial wave in three dimensions, or, if one additionally defines $V(r)=\infty$ for $r<0$, it can be thought of as a one-dimensional system. We follow previous analyses~\cite{winter61,garcia76,vandijk02} and choose the initial wave function of the particle as an eigenstate of the infinite square well with radius $a$,
\begin{equation}\label{eq:3.02}
\psi(r,0)=\phi_n(r)=\sqrt{\dfrac{2}{a}}\sin\left(\dfrac{n\pi r}{a}\right)\theta(a-r)
\end{equation}
where $n=1,2,\dots$ and $\theta(x)$ is the Heaviside function.  We obtain the exact wave function $\psi(r,\tau)$ for values of $r$ inside and outside the potential barrier; it is given by~\cite{vandijk02} 
\begin{equation}\label{eq:3.03}
\bs
\psi(r,\tau)=\sum_\nu & c_\nu\left\{{\cal M}(k_\nu, r-a,\tau) + \dfrac{i\lambda}{2k_\nu a}\theta(a-r)\right.  \\ 
& \left.\times \left[{\cal M}(k_\nu, r-a,\tau)-{\cal M}(k_nu,a-r,\tau)\right]\right.\Big\},
\end{split}
\end{equation}
where the $k_\nu$, $\nu =\pm 1,\pm 2,\dots$, are the solutions of the algebraic equation
\begin{equation}\label{eq:3.04}
ka \cot ka + \lambda - ika =0,
\end{equation}
and 
\begin{equation}\label{eq:3.05}   
c_\nu = \dfrac{(-1)^n 2n\pi\sqrt{2a}k_\nu}{(k_\nu^2a^2 - n^2\pi^2)
[(1+\lambda-ik_\nu a)\cot k_\nu a - i - k_\nu a]}.
\end{equation}
The function
\begin{equation}\label{eq:3.06}
{\cal M}(k,x,t) =   M(k,x,t) + \dfrac{1}{k}\chi(x,t) 
\end{equation} 
where $\ds \chi(x,t) = \dfrac{e^{\ts i\pi/4}}{2\sqrt{\pi t}}\exp\left(\dfrac{ix^2}{4t}\right)$,
and $M(k,x,t)$ is the Moshinsky function which for our purposes is defined as~\cite{vandijk02}
\begin{equation}\label{eq:3.07}
M(k,x,t) = \dfrac{1}{2} e^{\ts -ik^2t}e^{\ts ikx}\mathrm{erfc}(y), \ \ \ y=e^{\ts -i\pi/4}
\dfrac{x-2kt}{2\sqrt{t}}.
\end{equation}

\begin{figure}[h]
  \centering
   \resizebox{3.5in}{!}{\includegraphics[width=.5\textwidth, angle=-90]{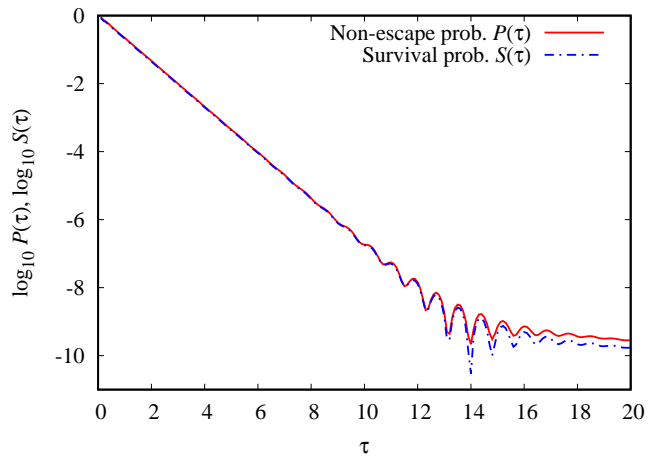}}
   \caption{Non-escape and survival probabilities as a function of time for $a=1$ and $\lambda=6$.
   The units of $\tau$ are generic as explained in the text.}
\label{fig_02}
\end{figure} 

The survival probability $S(\tau)$ and the non-escape probability $P(\tau)$ are, respectively,
\begin{equation}\label{eq:3.09}
\bs
S(\tau) & = \left|\int_{\ts 0}^{\ts\infty} dr\;\psi^*(r,0)\psi(r,\tau)\right|^2  \\
P(\tau) & = \int_{\ts 0}^{\ts a} dr\; |\psi(r,\tau)|^2. 
\end{split}
\end{equation} 
The probability density and  the probability current density are
\begin{equation}\label{eq:3.10}
\bs
\rho(r,\tau) & =\psi^*\psi \\
 j(r,\tau)  & = -i\left[\psi^*\dfrac{\partial\psi}{\partial r}-\dfrac{\partial\psi^*}{\partial r}\psi\right].
\end{split}
\end{equation}
These satisfy the equation of continuity,
\be{3.10a}
\dfrac{\partial~}{\partial\tau}\rho(r,\tau) + \dfrac{\partial~}{\partial r}j(r,\tau) = 0,
\ee
 which by integration over $r$ from zero to $a$ gives a relationship of the non-escape probability and the probability current density at $r=a$,  
\be{3.10a.1}
\dfrac{d~}{d\tau}P(\tau) = - j(a,\tau).
\ee

\begin{widetext}
\begin{figure}[!h]
$\begin{array}{cc}
\subfigure
{
\resizebox{1\linewidth}{!}{\includegraphics[width=0.5\textwidth, angle=-90]
{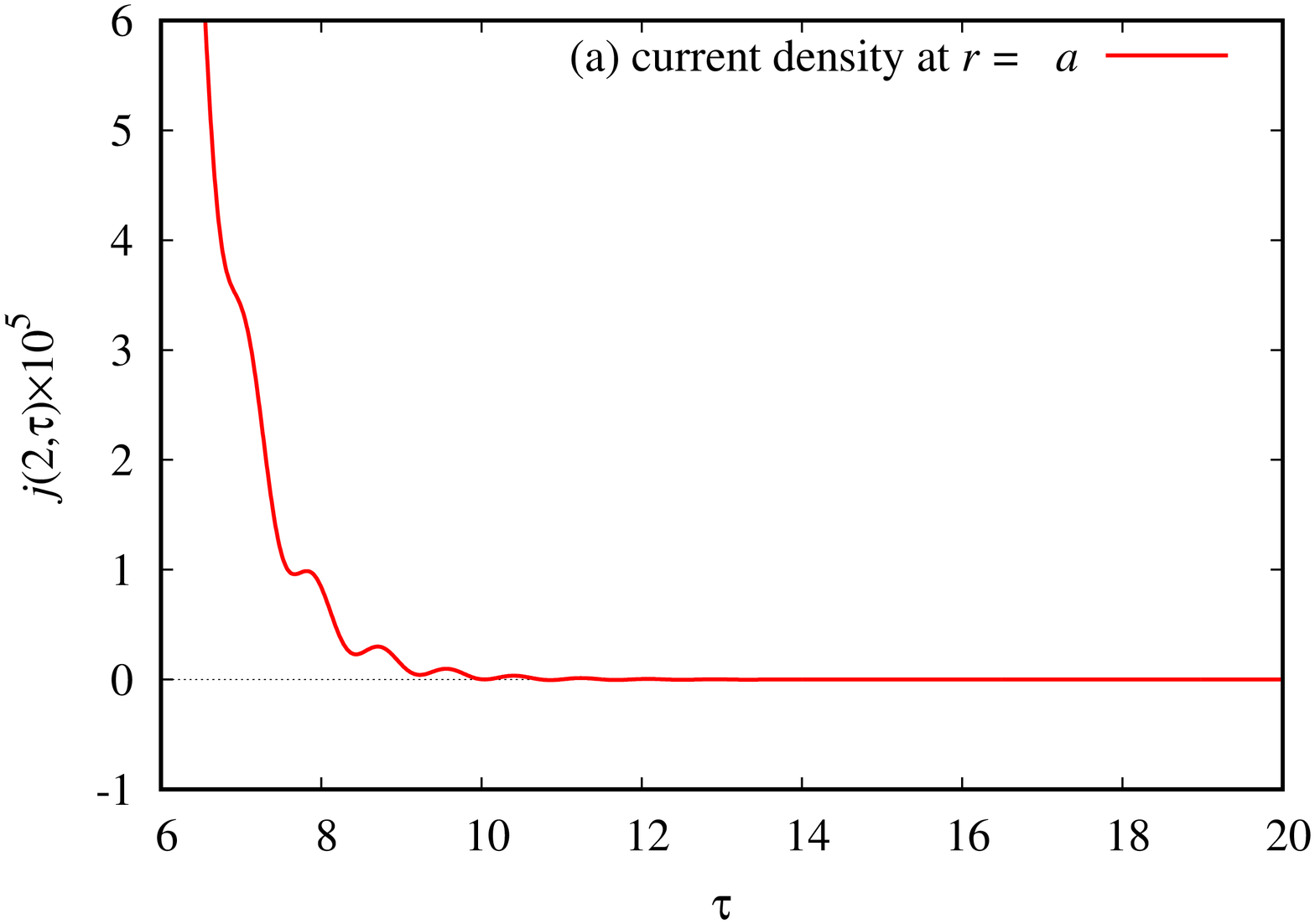}}
}
\label{fig:03a} &
\subfigure
{
\resizebox{1\linewidth}{!}{\includegraphics[width=.5\textwidth, angle=-90]
{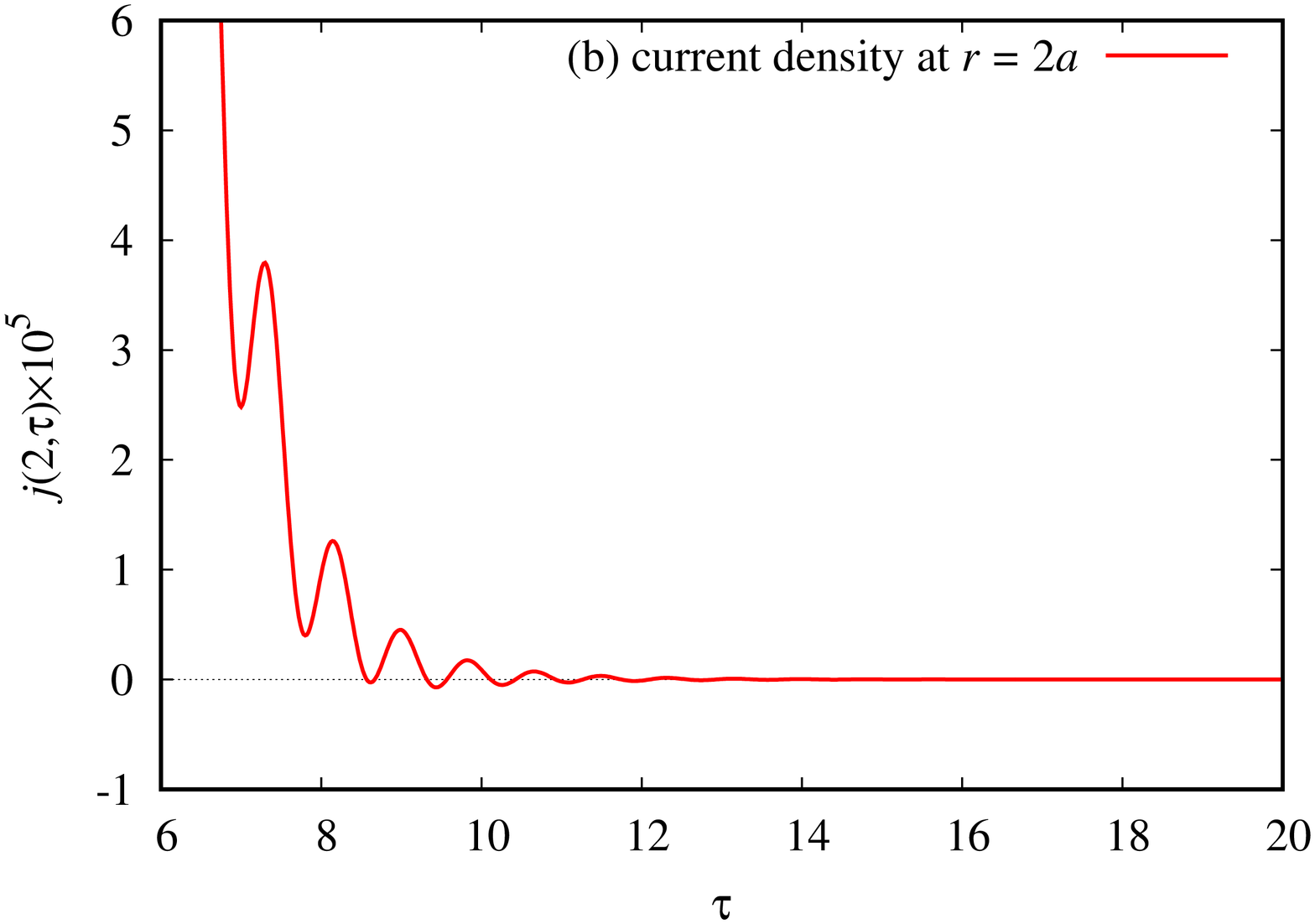}}
}
\label{fig:03b} \\
\subfigure
{
\resizebox{\linewidth}{!}{\includegraphics[width=.5\textwidth, angle=-90]
{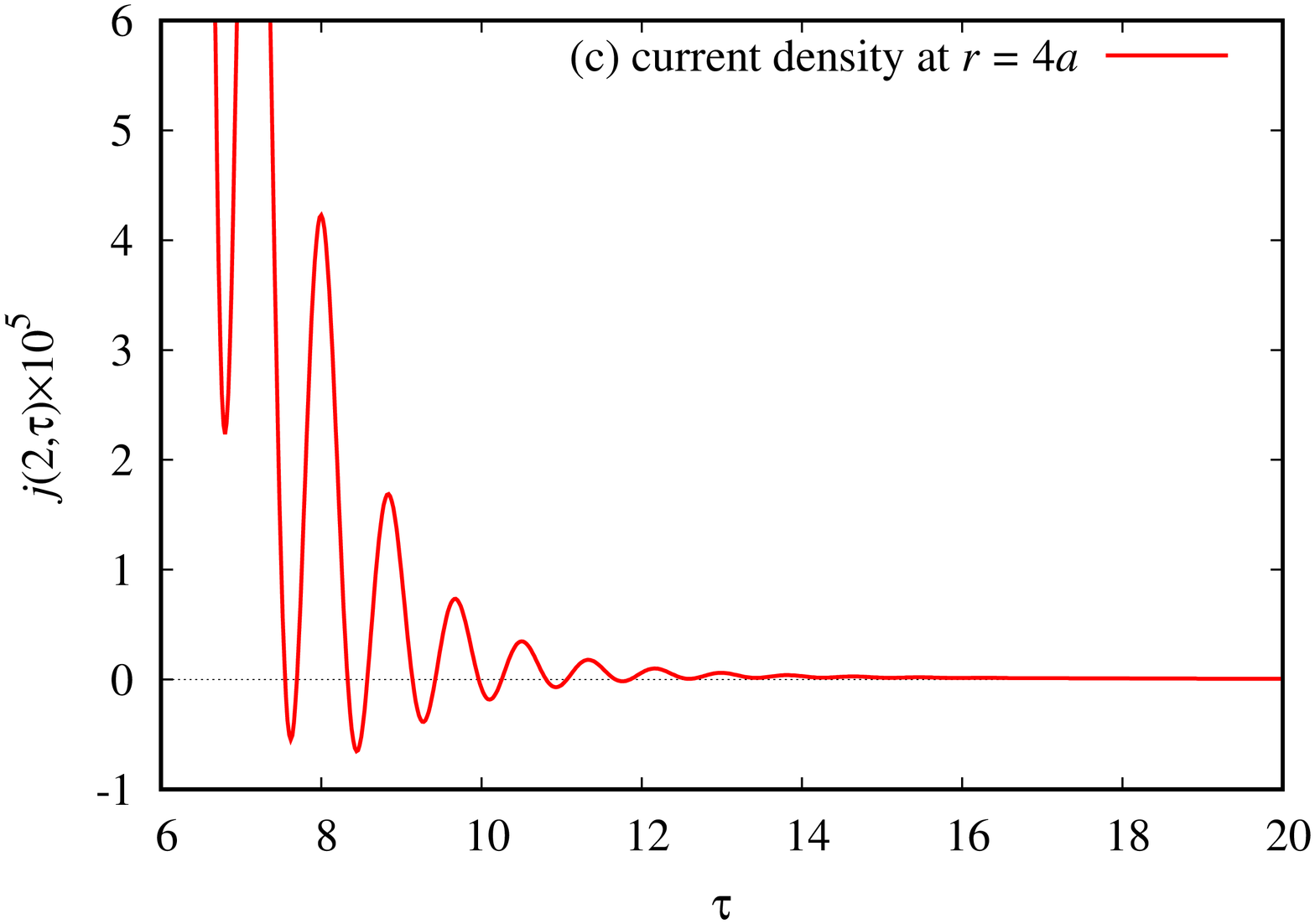}}
}
\label{fig:03c} &
\subfigure
{
\resizebox{\linewidth}{!}{\includegraphics[width=.5\textwidth, angle=-90]
{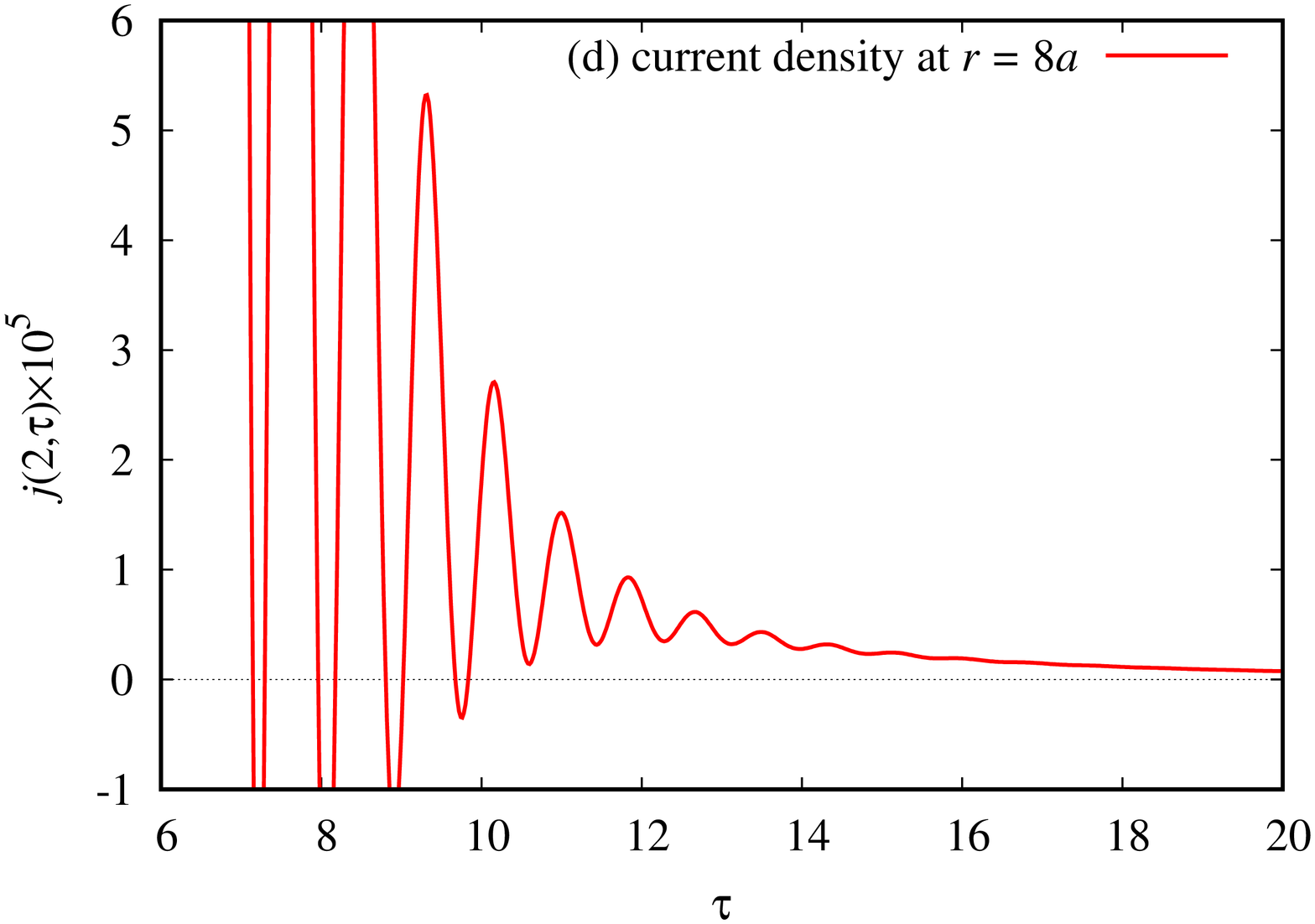}}
}
\label{fig:03d} \\ 
\end{array}$
\caption{\protect The current density at four different distances from the potential region when $a=1$ and $\lambda=6$.  The units of $a$ are arbitrary length units; the  units of $j$ are correspondingly inverse length units squared.}
\label{fig:03_total}
\end{figure}
\end{widetext}


The survival and non-escape probabilities as functions of time are plotted in Fig.~\ref{fig_02} for typical parameters, $\lambda = 6$ and $a=1$.  
The decay probabilities are not exponential at all times, but they fluctuate when the system transitions from the exponential decay to the long-time inverse power-law decay.  The fluctuations are significant since the temporary positive slopes indicate an increase, rather than a decrease, of the probability of finding the particle inside the potential barrier. 

 The probability current densities are plotted in Fig.~\ref{fig:03_total}\ifdraft{\footnote{Program: /home/vandijk/2014/TPBC/fortran/1p\_model\_1\_dp.f}}\fi ~at four different distances from the potential barrier, at $r=a$, $r=2a$, $r=4a$, and $r=8a$.  For the current one needs the spatial derivative of $\psi(r,\tau)$.  It is given\footnote{Unfortunately the factor in parentheses in the second term of Eq.~(\ref{eq:3.11}) is omitted in Ref.~\cite[Eq.~(43)]{vandijk02}.} for $0\leq r < \infty$ 
\begin{equation}\label{eq:3.11}
\bs
\dfrac{\partial~}{\partial r}\psi(r,t)  = & i\sum_\nu c_\nu\Big[k_\nu M(k_\nu,r-a,t) \ \\
& + \Big(1+\dfrac{r-a}{2tk_\nu}\Big)\chi(r-a,t)\Big] \\
& -\theta(a-r)\dfrac{\lambda}{2a}\sum_{\nu}c_\nu[M(k_\nu,r-a,t) \\
&+ M(k_\nu,a-r,t)+2\chi(r-a)/k_\nu].
\ \ \ \  \ifdraft{\checkmark}\fi
\end{split}
\end{equation}
Surprisingly the amplitude of the fluctuations of the probability current density increases the further out one goes. In fact the further out the more negative the current can be.

\begin{figure}[b]
\resizebox{1\linewidth}{!}{\includegraphics[width=.5\textwidth, angle=-90]
{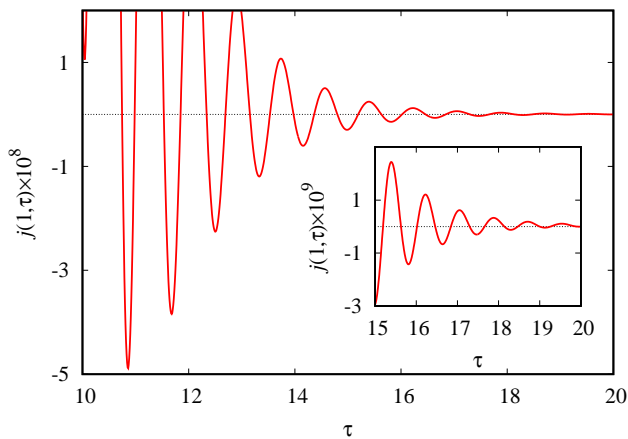}}
\caption{The enlarged profile of the probability density function $j(a,\tau)$ at the position of delta-function potential.  The times at which the current becomes negative are clearly seen. }
\label{fig:3.04} 
\end{figure}

A zoomed-in version\ifdraft{\footnote{$\sim$/backflow/calculations/backflow\_times/1p\_model\_1\_dp\_v3.f}}\fi ~of Fig.~\ref{fig:03_total}(a) is given in Fig.~\ref{fig:3.04}.
In this case there are twelve time intervals, $(\tau_{2i-1},\tau_{2i}), i=1,\dots,12$, during which $j(a,\tau)$ is negative.  Over each of these intervals we calculate the increase in probability of finding the particle to the left of the potential barrier,
\be{3.13}
\Delta_i = P(\tau_{2i})-P(\tau_{2i-1}) = - \int_{\ts\tau_{2i-1}}^{\ts\tau_{2i}}d\tau\; j(a,\tau).
\ee 
\begin{table}[h]
\caption{The time intervals $(\tau_{2i-1},\tau_{2i})$ during which the probability current density at the potential boundary $a$ is negative and the nonescape probability at $\tau_{2i-1}$ and the increase of the nonescape probability during the time interval.}\label{table:01}
\begin{tabular}{ccccc}
\hline
$i$ & \multicolumn{1}{c}{$~~~~~\tau_{2i-1}~~~~~$} & \multicolumn{1}{c}{$~~~~~~\tau_{2i}~~~~~~$}  & $P(\tau_{2i-1})$ & $\Delta_i$ \\ 
\hline\hline
1 &   10.745 &     10.983  & $~~~4.742\times 10^{-08}~~~$  & $~~~7.620\times 10^{-09}~~~$ \\     
2 &   11.532 &     11.847  & $1.074\times 10^{-08}$  & $7.824\times 10^{-09}$ \\     
3 &   12.338 &     12.693  & $1.979\times 10^{-09}$  & $5.149\times 10^{-09}$ \\     
4   & 13.154 &     13.531  & $3.116\times 10^{-10}$  & $2.887\times 10^{-09}$ \\     
5  & 13.975 &      14.364  & $1.978\times 10^{-10}$  & $1.498\times 10^{-09}$ \\     
6   & 14.801 &     15.193  & $3.050\times 10^{-10}$  & $7.420\times 10^{-10}$  \\     
7  & 15.629 &      16.019  & $3.755\times 10^{-10}$  & $3.545\times 10^{-10}$ \\     
8  & 16.461 &      16.841  & $3.927\times 10^{-10}$  & $1.624\times 10^{-10}$ \\     
9   & 17.298 &     17.659  & $3.777\times 10^{-10}$  & $6.948\times 10^{-11}$ \\     
10  & 18.142 &    18.467  & $3.482\times 10^{-10}$  & $2.595\times 10^{-11}$ \\     
11  & 19.001 &    19.259  & $3.143\times 10^{-10}$  & $6.804\times 10^{-12}$ \\     
12  & 19.912 &    19.998  & $2.804\times 10^{-10}$  & $1.364\times 10^{-13}$ \\ \hline
\end{tabular}
\end{table}    
Table~\ref{table:01} lists the times at which the probability current density is zero and the increase of the probability of the particle inside the potential which occurs when the current is negative. \ifdraft{\footnote{$\sim$/2019/backflow/calculations/backflow\_times/1p\_model\_1\_dp\_v2.f}.}\fi      ~We also list the nonescape probability at the beginning of each interval.  The fluctuations in the current occur a long time into the decay and hence the nonescape probability is already very small.    Nevertheless $\Delta_i$ can be significantly larger than $P(\tau_{2i-1})$.  In other words the probability increase during the time interval of negative probability current density can exceed the nonescape probability at the beginning of the time interval.  

We also study the probability current densities at $r=8a$ (Fig.~\ref{fig:03_total}(d)), where remarkably the negative amplitudes are much larger than at $r=a$ although, in the case shown, there are fewer time intervals with negative probability current density.  The intervals are listed in Table~\ref{table:02}.  

\begin{table}[h]
\caption{The time intervals $(\tau_{2i-1},\tau_{2i})$ during which the probability current density at $r=8a$ is negative, and the nonescape probability at $\tau_{2i-1}$ (here defined as probability of finding the particle in the region $r\in(0,8a)$), and the increase of this nonescape probability during the time interval.}\label{table:02}
\begin{tabular}{ccccc}
\hline
$i$ & \multicolumn{1}{c}{$~~~~~\tau_{2i-1}~~~~~$} & \multicolumn{1}{c}{$~~~~~~\tau_{2i}~~~~~~$}  & $P_{r<8a}(\tau_{2i-1})$ & $\Delta_i$ \\ 
\hline\hline
1 &   7.146 &     7.295  & $~~~2.079\times 10^{-4}~~~$  & $~~~3.465\times 10^{-6}~~~$ \\     
2 &   7.963 &     8.177  & $1.022\times 10^{-4}$  & $4.367\times 10^{-6}$ \\     
3 &   8.806 &     9.023  & $6.351\times 10^{-5}$  & $2.015\times 10^{-6}$ \\     
4 &   9.675 &     9.835  & $4.543\times 10^{-5}$  & $3.689\times 10^{-7}$ \\     
\hline
\end{tabular}
\end{table}    

\section{Quantum backflow interpretation}
\label{sec:03}
\begin{figure}[h]
\resizebox{1\linewidth}{!}{\includegraphics[width=.5\textwidth, angle=-90]
{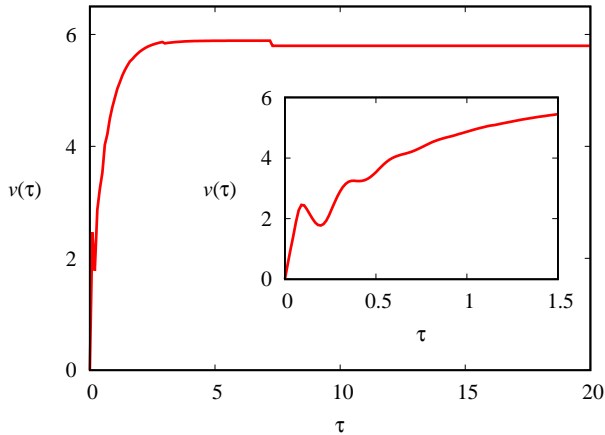}}
\caption{The mean velocity of the particle as a function of time for $a=1$ and $\lambda=6$.  The units of velocity are inverse length.}
\label{fig:08} 
\end{figure}

\setcounter{equation}{0}
From a classical point of view one expects that particles emitted from a decaying system travel away from the source.  It is therefore surprising that the quantum probability current density at times is negative or inward, even at the potential barrier.  Consider the expectation value of the position and velocity of the particle as functions of time
\be{bi:01}
\bs
x(\tau) & = \int_{\ts 0}^{\ts\infty} dr\; \psi^*(r,\tau) r \psi(r,\tau) \\
v(\tau) & =  \int_{\ts 0}^{\ts\infty} dr\; \psi^*(r,\tau) \left[(-i)\dfrac{\partial~}{\partial r}\psi(r,\tau)\right].
\end{split}
\ee    
In Fig.~\ref{fig:08} the mean velocity is plotted as a function of time\ifdraft{\footnote{$\sim$/2019/backflow/calculations/backflow\_times/1p\_model\_1\_dp\_v5.f}}
\fi 
with the structure of the graph for small $\tau$ shown in the inset. The mean velocity is nonnegative at all times becoming nearly constant for large times. We note that the mean velocity is outward at all times even during the time intervals when the probability current density is negative.  According to the Ehrenfest theorem
\be{bi:01a}
\dfrac{d~}{d\tau} x(\tau) = v(\tau),
\ee
so that $x(\tau)$ is monotonically increasing with time.
 
\begin{figure}[h]
\resizebox{1\linewidth}{!}{\includegraphics[width=.5\textwidth, angle=-90]
{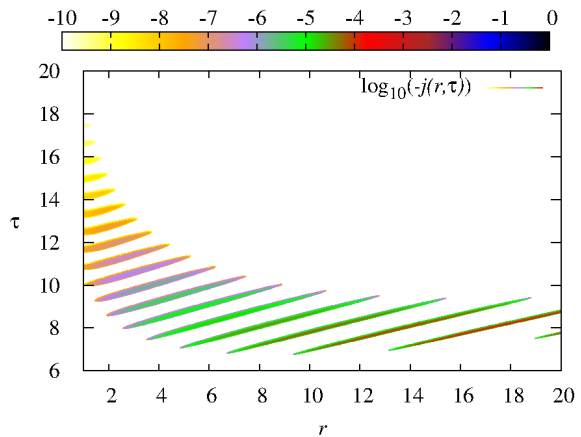}}
\caption{The regions coloured (or grey shaded) on the $r\tau$ plane  where the probability current density is negative for $a=1$ and $\lambda=6$.}
\label{fig:10} 
\end{figure}

It is instructive to plot the regions of negative probability current density on the $r\tau$ plane~\cite{berry10} as in Fig.~\ref{fig:10}.  In the graph the magnitude of the $\log_{10}[-j(r,\tau)]$ is plotted as a function of $(r,\tau)$ according to the color code indicated.  Points at which the probability current density is positive are left blank.  The graph provides a pictorial display of the ``islands of backflow,"  and shows succinctly and generally the salient features of Figs.~\ref{fig:03_total} and \ref{fig:3.04}.   
The somewhat regular pattern shows a periodic behaviour at the potential boundary, $r=a$, where the probability flows in and out of the potential region.  This is also vividly demonstrated using Bohmian quantum trajectories~\cite{nogami00}.  We note that the graph depends on having exact wave functions for $r>a$. 

The discussions of quantum backflow~\cite{allcock69c,bracken94,yearsley13,berry10,yearsley13} focus mainly on free particles traversing the origin from left to right in one-dimensional space, whose wave functions in momentum space have zero amplitude for negative momentum components.  Quantum backflow occurs when the probability of the particle being on the left side of the origin increases temporarily.  In our system the particle is initially confined to a region in coordinate space, but escapes in time by tunnelling though the barrier.  However, again there are times that the probability of the particle being inside the potential region increases temporarily.  So we identify this also as quantum backflow.  
Actually this phenomenon is closer to quantum reentry discussed by Goussev~\cite{goussev19}, who determined that quantum backflow and quantum reentry are equivalent.   The phenomenon is general in the sense that it is not an artifact of the nature of the delta-function barrier.  Exact solutions for such systems with bound states and/or with square barriers~\cite{vandijk02} or numerical solutions for Gaussian barriers~\cite{vandijk07} all show similar behaviour.

The experimental observation of the nonexponenial decay is discussed recently by Ram\'irez Jim\'enez and Kelkar~\cite{ramirez_jimenez19}.  Their conclusion is that the nonexponential decay is unlikely to be seen in unstable nuclei and particles.  However, experimental evidence for nonexponential luminescence decay of excited dissolved organic materials is reported in Ref.~\cite{rothe06}.  The experiments are not sufficiently precise to detect the fluctuation in the transition regions.  However the recent experiments of Crespi \etal~\cite{crespi19} using integrated photonics do show \mbox{short-,} intermediate-, and long-time effects of quantum decay including the oscillatory behaviour between the exponential and long-time inverse-power law modes of behaviour.  

\begin{figure}[h]
\centering
\resizebox{1\linewidth}{!}{\includegraphics[width=.5\textwidth, angle=-90]
{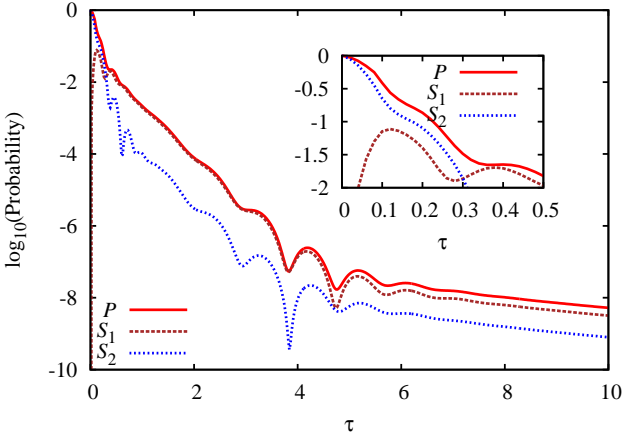}}
\caption{The nonescape and survival probabilities of a system initially in state $\phi_2(r)$.  The quantity $S_1(\tau)$ is the probability of finding the system in state $\phi_1(r)$ at time $\tau$.  The parameters of the calculation are $n=2$, $\lambda = 3$ and $a=1$.}
\label{fig:11} 
\end{figure}

The backflow is related to the oscillatory behaviour of the probability current density occurring in the transition region from exponential decay to 
inverse-power law decay.  Ram\'irez Jim\'enez and Kelkar~\cite{ramirez_jimenez19} suggest that the survival probability in the transition period is obtained by combining the survival amplitude of the exponential decay and the survival amplitude of the inverse-power law decay.   This leads to an interference term in the survival probability which shows up as oscillations.  It may be of interest if such interference also occurs when one transitions from one exponential decay region to another.  In Fig.~\ref{fig:11} we show the nonescape probability\ifdraft{\footnote{$\sim$/2019/backflow/calculations/nonescape/1p\_model\_1\_dp\_v6.f}}\fi ~when the initial state is characterized with $n=2$.  In this case we define survival probabilities $S_n(\tau)$
\be{4.02}
S_n(\tau) = \left|\int_{\ts 0}^{\ts a} dr \; \phi_n^*(r)\psi(r,\tau)\right|^2.
\ee 
Since the initial state has $n=2$,  $S_2(\tau)$ is the true survival probability, and  $S_1(\tau)$ is the probability of finding the system in state $n=1$ at time $\tau$.  Initially $S_1(\tau)$ is zero but as the system evolves in time the $n=2$ state is depleted and the $n=1$ state builds up and decays at a slower rate.
We note in Fig.~\ref{fig:11} that the nonescape probability increases by a small, but significant, amount from time $\tau=0.36$ to $\tau=0.40$.  Hence we observe backflow during the transition from one type of exponential decay to another.  

  As mentioned, most of the earlier backflow studies  involve quantum wave packets with positive wave numbers passing some point on the one-dimensional line, usually the origin.  In the model we study the initial wave function has momentum components in both direction.  But the direction of the negative momentum wave is reversed as it reflects from the infinite barrier and is positive as it arrives at the potential barrier at $r=a$.  The backflow occurs when there is complicated interference of the immediate incident and reflected waves.
   
\section{``Decay" of free wave packet}
\label{sec:04}
\setcounter{equation}{0}

\begin{figure}[h]
\centering
\resizebox{1\linewidth}{!}{\includegraphics[width=.5\textwidth, angle=-90]
{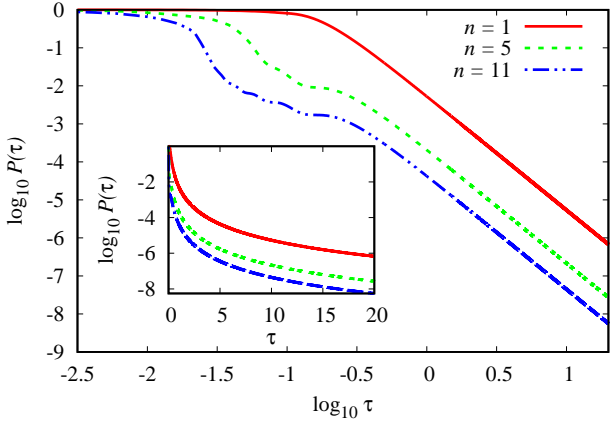}}
\caption{The nonescape probability of a free particle originally located in region $(0,a)$ when $a=1$ and $n=1$, 5, and 11.}
\label{fig:12} 
\end{figure}

Since backflow is purported to be a universal quantum effect, existing in an interaction-free environment as well as in the presence of short-range potentials~\cite{bostelmann17}, we further elucidate the decaying behaviour of the preceding section by examining the time evolution of a free wave packet initially confined to the region $r\in(0,a)$.  To do so we first need the $S$-wave partial-wave propagator of a free particle.  For an interaction-free system the time-independent eigenstates  that vanish at the origin are
\be{5.01}
\psi_k(r) = \sqrt{\dfrac{2}{\pi}}\sin(kr) \ \ \ \ \mathrm{for} \  \ 0\le k<\infty, \ \ 0\le r < \infty.
\ee
Normalization, orthogonality and completeness result in the following conditions, 
\be{5.02}
\bs
\int_0^\infty dk \; \psi_k(r)\psi_k(r') & = \delta(r-r'), \\
 \int_0^\infty dr \; \psi_k(r) \psi_{k'}(r) & = \delta(k-k').
 \end{split}
\ee
Furthermore, in terms of the Hamiltonian $\widehat{H}$,
\be{5.03}
\widehat{H}\psi_k= \left(-\dfrac{\hbar^2}{2m}\dfrac{\partial^2~}{\partial r^2}\right)\psi_k= \dfrac{\hbar^2k^2}{2m}\psi_k = \epsilon_k \psi_k.
\ee
The free-particle propagator is \ifdraft{\footnote{$\sim$/2019/masa/May\_02\_free.mw}}\fi
\be{5.04}
\bs
K(r,t;r',t_0)& = \int_0^\infty dk \; \psi_k(r)e^{\ts -i\widehat{H}(t-t_0)/\hbar}\psi_k^*(r') \\
& = \dfrac{2}{\pi}\int_0^\infty dk \sin(kr)e^{\ts -i \epsilon_k(t-t_0)/\hbar} \sin(kr') \\
& = \dfrac{e^{\ts 3i\pi/4}}{\sqrt{2\pi\tau}}\left[e^{\ts i(r+r')^2/(4\tau)} \right. \\ 
& \left. \hspace{1in} - e^{\ts i(r-r')^2/(4\tau)}\right], \ifdraft{\checkmark}\fi
\end{split}
\ee
where again we define $\tau=\hbar(t-t_0)/(2m)$.
The wave function for any $\tau$ can be calculated using
\be{5.05}
\psi(r,\tau) = \int_0^\infty dr' \; K(r,t;r',t_0)\psi(r',0).
\ee
Consider now the initial wave function (localized as with the $\delta$-shell potential)
\be{5.06}
\psi_n(r,0) = \sqrt{\dfrac{2}{a}}\sin\left(\dfrac{n\pi r}{a}\right)\theta(a-r)
\ee
with $n$ a positive integer.
The time-dependent wave function is 
\be{5.07}
\bs
\psi_n(r,\tau) = & -\dfrac{i}{2\sqrt{2a}}e^{\ts -i\pi^2n^2\tau/a^2}\left(\left\{\mathrm{erf}[\xi_n^{(+)}(r,-\tau)/\sqrt{\tau}] \right.\right.  \\
& \hspace{-.3in}\left. +\mathrm{erf}[\xi_n^{(-)}(r,\tau)/\sqrt{\tau}]\right\} 
 e^{\ts i\pi nr/a} - \left\{\mathrm{erf}[\xi_n^{(+)}(r,\tau)/\sqrt{\tau}] \right. \\
& \hspace{-.3in}\left.\left. + \mathrm{erf}[\xi_n^{(-)}(r,-\tau)/\sqrt{\tau}]\right\}e^{\ts -i\pi nr/a}\right),  \ifdraft{\checkmark}\fi
\end{split}
\ee
where
\be{5.08}
\xi_n^{(\pm)}(r,\tau) = (1-i)(2\pi n\tau + a^2 \pm ar)/(2a\sqrt{2}).
\ee
By invoking the relationship\ifdraft{\footnote{See MAPLE.}}\fi~~ $\ds\lim_{x\rightarrow\pm\infty}\left[\mathrm{erf}\left(e^{-3i\pi/4}x\right)\right] = \mp 1$, it is straightforward to show that $\psi_n(r,\tau)$ reduces to $\psi_n(r,0)$ of Eq.~(\ref{eq:5.06}) as $\tau$ approaches $0$.
Using the wave function of Eq.~(\ref{eq:5.07}), we calculate the nonescape probability as a function of time\ifdraft{\footnote{$\sim$/2019/backflow/calculations/free\_nonescape/free\_nonescape\_v2a.f}}\fi, and show the result in Fig.~\ref{fig:12}.
Typical features of decaying systems are in evidence in this free-particle situation: the quadratic-time behaviour at short time, followed by exponential decay, somewhat erratic transition to a clean $\tau^{-3}$ long-time fall off.  The last property follows from the long-time dependence of the propagator\ifdraft{\footnote{$\sim$/2019/masa/May\_02\_free.mw}}\fi,
\be{5.09} 
K(r,t;r',t_0) \sim \dfrac{e^{\ts 3i\pi/4}}{2\sqrt{\pi}}rr'\tau^{-3/2} +{\cal O}(\tau^{-5/2}),
\ee
which yields the wave function at large $\tau$
\be{5.10}
\psi_n(r,\tau) \sim  \dfrac{e^{\ts i\pi/4}}{\sqrt{2}}\left(\dfrac{a}{\pi}\right)^{3/2}\dfrac{(-1)^n}{n}r\tau^{-3/2} 
+ {\cal O}(\tau^{-5/2}).
\ee
Such a wave function is useful in calculating the nonescape probability at large times.  Actually the long-time behaviour of the survival probabilities of free particles is discussed by Miyamoto~\cite{miyamoto02}; it varies as $\tau^{-(2\ell+1)}$, where $\ell$ characterizes the small $k$ behaviour of the Fourier transform of the initial wave function as constant$\times k^\ell$.

\begin{figure}[h]
\centering
\resizebox{1\linewidth}{!}{\includegraphics[width=.5\textwidth, angle=-90]
{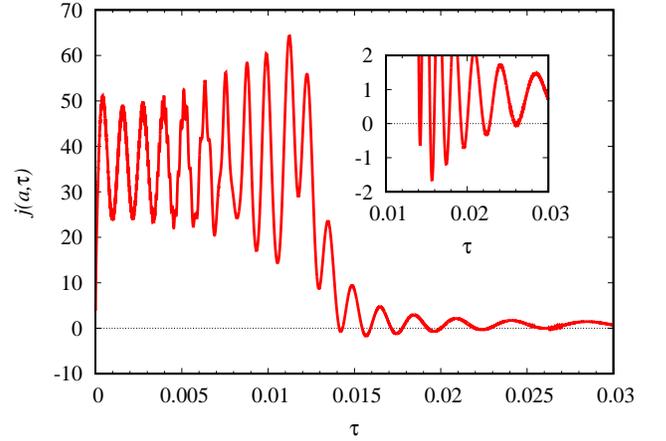}}
\caption{The probability current density at $r=a=1$ as a function of time for the free particle for $\Psi(r,\tau) = 1/\sqrt{2}[\psi_1(r,\tau) + e^{\ts i\pi/4}\psi_{23}(r,\tau)]$. }
\label{fig:14} 
\end{figure}

For the calculation of the current we need the spatial derivative of the wave function,
\be{5.11}
\bs
&\dfrac{\partial\psi_n}{\partial r}(r,\tau) =  \dfrac{n\pi}{2\sqrt{2}a^{3/2}}e^{\ts -i\pi^2n^2\tau/a^2} \\
& \times \left\{\left(\mathrm{erf}\left[\dfrac{\xi_n^{(+)}(r,-\tau)}{\sqrt{\tau}}\right] + \mathrm{erf}\left[\dfrac{\xi_n^{(-)}(r,\tau)}{\sqrt{\tau}}\right]\right)e^{\ts i\pi nr/a} \right.\\
& \left.+\left(\mathrm{erf}\left[\dfrac{\xi_n^{(+)}(r,\tau)}{\sqrt{\tau}}\right] +  \mathrm{erf}\left[\dfrac{\xi_n^{(-)}(r,-\tau)}{\sqrt{\tau}}\right]\right)e^{\ts -i\pi nr/a}\right\}.
\end{split}
\ee
Since the $\psi_n(r,0)$ form a complete set of states (they are eigenstates of the infinite square well) we can, by superposition,  start with an initial state of any shape as long as it is zero at $r=0$ and $r=a$.  Thus
\be{5.12}
\Psi(r,0) = \sum_{n=1}^\infty c_n \psi_n(r,0) \ \ \ \mathrm{with} \ \ \ \sum_{n=1}^\infty |c_n|^2 = 1.
\ee
Then the wave function and its spatial derivative at any later time are
\be{5.13}
\Psi(r,\tau) = \sum_{n=1}^\infty c_n \psi_n(r,\tau)  \ \ \mathrm{and}  \ \ \dfrac{\partial\Psi}{\partial r}(r,\tau) = \sum_{n=1}^\infty c_n \dfrac{\partial\psi_n}{\partial r}(r,\tau).
\ee
In Fig.~\ref{fig:14} \ifdraft{\Red $\checkmark$}\fi   we plot the probability current density as a function of time\ifdraft{\footnote{ $\sim$/2019/backflow/calculations/free\_nonescape/free\_nonescape\_v4a\_new.f.}}\fi .
We use a typical (and arbitrary) combination of  $n=1$ and $n=23$ waves, $\Psi(r,\tau) =[\psi_1(r,\tau)+e^{\ts i\pi/4}\psi_{23}(r,\tau)]/\sqrt{2}$.   We observe the following.
\begin{enumerate}
\item There are at least five, perhaps six, time intervals during which the probability density current is negative.  
\item At early times the probability current density is positive.  This is indicative of the fact that at $\tau=0$ the wave functions consist of components with positive and negative wave numbers.  The latter travel to the left initially and are reflected at the origin.  They combine with the initially right moving components so that all components crossing the $r=a$ point are moving to the right.  The negative probability current density occurs at later times when significant interference can occur.    
\item The backflow occurs when there is interference of wave functions components with different values of $n$. 
\end{enumerate}

\begin{figure}[h]
\centering
\resizebox{1\linewidth}{!}{\includegraphics[width=.5\textwidth, angle=-90]
{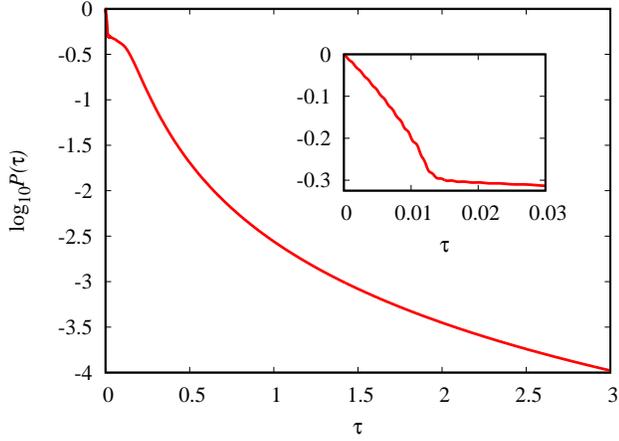}}
\caption{The logarithm of the nonescape probability of the free particle as a function of time  for $a=1$ and the parameters of Fig.~\ref{fig:14}.}
\label{fig:15} 
\end{figure}

\begin{figure}[h]
\centering
\resizebox{1\linewidth}{!}{\includegraphics[width=.5\textwidth, angle=-90]
{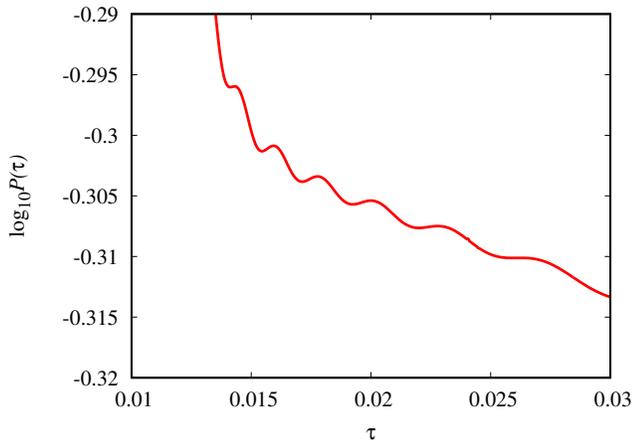}}
\caption{The logarithm of the nonescape probability of the free particle as a function of very short times  for $a=1$ and the parameters of Fig.~\ref{fig:14}.}
\label{fig:16} 
\end{figure}

It is enlightening to relate the negative probability current density intervals to the nonescape probability shown\ifdraft{\footnote{$\sim$/2019/backflow/calculations/free\_nonescape/free\_nonescape\_v4b\_new.f}}\fi   ~in Fig.~\ref{fig:15}.   
Given that the energy of the $n=23$ component is much larger than that of the $n=1$ component, the former decays much faster.  This results in the precipitous drop of the escape probability at very short times.  Once the $n=23$ component is nearly depleted the $n=1$ component continues to decrease according to its rate of decay.  This leads to an abrupt change in the slope of the $P$ versus $\tau$ curve. 
The decay continues according the exponential decay of the component $n=1$ until at long times the nonescape probability attribute converts to a $\tau^{-3}$ behaviour.  At the changeover of the $n=23$ exponential decay to the $n=1$ exponential decay we note fluctuations in the escape probability.  Figure~\ref{fig:16} shows an expanded view  of the nonescape probability as a function of a time over a range involving only very short times.
It is noteworthy that the nonescape probability increases over the time intervals during which the probability current density is negative.  Compare Figs.~\ref{fig:14} and \ref{fig:16}.  This occurs when the decay process transitions from one decay mode ($n=23$) to another ($n=1$).  At this point we do not see similar fluctuations when the decay rate changes from the exponential to the inverse power law in time.
\section{Backflow as eigenvalue problem}
\label{sec:05}
\setcounter{equation}{0}
 Bracken and Melloy~\cite{bracken94}  maximize the increase of backflow probability in their model by means of an eigenvalue method.   They investigate the evolution of an initial wave function consisting of components with nonnegative wave numbers, and  calculate the probability $P(t)$ of the particle in the region  $x\in(-\infty,0)$ as a function of time.  Backflow is indicated if there are time intervals during which $P(t)$ increases.  The maximum backflow they obtain is independent of any dimensioned quantity (e.g., mass, $\hbar$, or the length of time that backflow occurs.)    There is however an upper limit on the amount of backflow during any given time interval, and it is found to be less than 4\%.   

The independence of the maximum backflow on the time interval is based on a scaling property of the probability current density,
\be{6.00}
\tilde{j}(x,t) = \dfrac{1}{\mu^2}j\left(\dfrac{x}{\mu},\dfrac{t}{\mu^2}\right).
\ee
The backflow that occurs in the Bracken and Melloy model depends on the current at the origin which does not change with the scaling.  
In the case of the partial $S$ wave of a free particle, the particle is initially confined to the region $r\in(0,a)$ and the nonescape probability as a function of time is determined at some nonzero position, say $a$.   Since the scaling would give different positions, it is not meaningful to compare currents, and the amount of backflow cannot be expressed without reference to $a$ or time $\tau$.  Nevertheless we calculate a typical backflow and determine whether it is of the same order of magnitude as the dimensionless quantum number $c_{mb}  = 0.0384517\dots$~\cite{penz06} of Bracken and Melloy..

\begin{figure}[h]
\centering
\resizebox{1\linewidth}{!}{\includegraphics[width=.5\textwidth, angle=-90]
{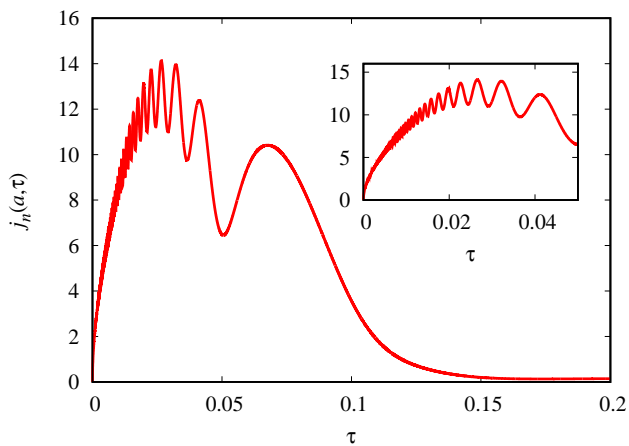}}
\caption{The probability current density $j_n(a,\tau)$ for small $\tau$ when $n=3$ and $a=1$.}
\label{fig:17} 
\end{figure}

  Using the results of the last section, the general time-evolving wave function of a free particle is
\be{6.01}
\psi(r,\tau)=\sum_{n=1}^\infty c_n\psi_n(r,\tau).
\ee
The probability current density is
\be{6.02}
j(r,\tau) 
=-i\sum_{n=1}^\infty\sum_{n'=1}^\infty c_n^*c_{n'}\left(\psi_n^*\dfrac{\partial\psi_{n'}}{\partial r}-\dfrac{\partial\psi_n^*}{\partial r}\psi_{n'}\right).
\ee
The backflow probability in the time interval $(\tau_l,\tau_u)$ is
\be{6.03}
\bs
\Delta_P & = -\int_{\ts\tau_l}^{\ts\tau_u} d\tau\; j(a,\tau) \\
&=i\sum_{n=1}^\infty\sum_{n'=1}^\infty c_n^*c_{n'}  \int_{\ts\tau_l}^{\ts\tau_u} d\tau\; \left(\psi_n^*\dfrac{\partial\psi_{n'}}{\partial r}-\dfrac{\partial\psi_n^*}{\partial r}\psi_{n'}\right) \\
& = \mathbf{c}^\dagger M \mathbf{c},
\end{split}
\ee
where $\mathbf{c} = (c_1,c_2,\dots,c_n,\dots)^T$, and
\be{6.04}
M_{nn'} = i\int_{\ts\tau_l}^{\ts\tau_u} d\tau\;\left[\psi_n^*\dfrac{\partial\psi_{n'}}{\partial r}
-\dfrac{\partial\psi_n^*}{\partial r}\psi_{n'}\right].
\ee
We determine the extrema of $\mathbf{c}^\dagger M\mathbf{c}$ with the constraint $\mathbf{c}^\dagger \mathbf{c} = 1$, using a Lagrange multiplier, i.e.,
\be{6.05}
\bs
I(\mathbf{c})& =\mathbf{c}^\dagger M\mathbf{c} - \lambda(\mathbf{c}^\dagger\mathbf{c}-1)\\
&= \sum_n\sum_{n'} c_n^*M_{nn'}c_{n'}-\lambda\sum_n c_n^*c_n +\lambda.
\end{split}
\ee
The extremum condition $\dfrac{\partial I}{\partial c^*_i} = 0$ for $i=1,2,\cdots$ leads to the eigenvalue equation
\be{6.06}
M\mathbf{c} = \lambda \mathbf{c}.
\ee
\ifdraft{(A more precise, and believable, derivation is given in Appendix~\ref{appen:b}.)\fi  ~Since the matrix $M$ is Hermitian the eigenvalues are real. 
Although the matrix elements of $M$ need to be evaluated only once, the integrals may present challenges since the integrand has oscillations whose frequency goes to infinity as $\tau$ approaches zero.  Figure~\ref{fig:17} illustrates the behaviour of $j_3(a,\tau)$ which is basically the integrand of $M_{nn}$  for \ifdraft{\footnote{$\sim$/2019/backflow/calculations/eigenvalue\_v4.f, ../jn2.gp}}\fi  ~$n = 3$.

To overcome the difficulties of evaluating the integrals in Eq.~(\ref{eq:6.04}) we make a substitution $\tau=1/z$, so that
\be{6.07}
\int_{\ts\tau_l}^{\ts\tau_u} d\tau\; f(\tau) = \int_{\ts 1/\tau_u}^{\ts 1/\tau_l} \dfrac{dz}{z^2}\; f\left(\dfrac{1}{z}\right) .
\ee
  \ifdraft{ This approach results in  accurate integrals.   (The approach is suggested by Cheney and Kincaid~\cite[page 243, problem 9]{cheney08} for a function $f(t)= t|\sin(1/t)|$ which has increasing frequency as $t\rightarrow 0$.)}\fi   ~The Romberg integration algorithm is efficient in yielding accurate integrals. 
We expect $-1\le\Delta_P\le1$, where $\Delta_P=-1$ is the case of no quantum backflow.  In fact $\Delta_P>0$ indicates a net backflow during the interval $(\tau_l,\tau_u)$.  \ifdraft{It was pointed out~\cite{bracken94,penz06,yearsley12,yearsley13} in the Bracken-Melloy analysis that $\Delta_P\le c_{mb} = 0.0384517\dots$~\cite{penz06}.  Whether this is also the case for our model remains to be seen.}\fi
 
Considering the time interval such that $\tau_l=0.02$ and $\tau_u=0.04$ we use the shifted power method algorithm to obtain the least (i.e., most negative) eigenvalue and the shifted inverse power method to obtain the largest eigenvalue\ifdraft{\footnote{$\sim$/2019/backflow/calculations/eigenvalue\_v7b.f}}\fi.  We do this for the various  maximum values of $n$ in the sums listed in Table~\ref{table:03}.
\begin{table}[!h]
\caption{Smallest and largest eigenvalues when $\tau_l=0.02$ and $\tau_u=0.04$\ifdraft{\footnote{$\sim$/2019/backflow/calculations/eigenvalue\_v7e.f (quadruple precision)}}\fi.  The quantity $n$ refers to the maximum value of $n$ in the sum of Eq.~(\ref{eq:6.01}).}
\label{table:03}
\begin{tabular}{ccccc}
\hline
~~$n$~~ & ~~~~~~~~$\lambda^{(l)}$~~~~~~~~ &~~~$e_1^{(l)}$~~&~~~~~~~~$\lambda^{(h)}$~~~~~~~~&~~~$e_1^{(h)}$~~~ \\
\hline\hline
2   &  $-$0.16424  & $1.47\times 10^{-7}$ & 0.00032 & $4.88\times 10^{-14}$\\
3   &  $-$0.40406  & $4.05\times 10^{-7}$ & 0.00093 & $2.37\times 10^{-12}$\\
4   &  $-$0.59755  & $5.77\times 10^{-11}$ & 0.00316 & $1.52\times 10^{-12}$\\
5   &  $-$0.79816  & $1.10\times 10^{-10}$ & 0.01143 & $3.76\times 10^{-13}$\\
6   &  $-$0.92258  & $5.17\times 10^{-8}$ & 0.01156 & $3.62\times 10^{-13}$\\
7   &  $-$0.98060  & $5.37\times 10^{-7}$ & 0.01525 & $2.71\times 10^{-13}$\\
8   &  $-$0.99419  & $6.44\times 10^{-7}$ & 0.01553 & $2.69\times 10^{-13}$\\
9   &  $-$0.99762  & $3.75\times 10^{-7}$ & 0.01743 & $2.43\times 10^{-13}$\\
10  &  $-$0.99901  & $3.21\times 10^{-7}$ & 0.01793 & $2.38\times 10^{-13}$\\
15  &  $-$0.99997  & $2.57\times 10^{-7}$ & 0.02287 & $1.72\times 10^{-13}$\\
20  &  $-$0.99998  & $2.55\times 10^{-7}$ & 0.02409 & $1.64\times 10^{-13}$\\
25  &  $-$0.99998  & $2.55\times 10^{-7}$ & 0.02457 & $1.58\times 10^{-13}$\\ 
50  &  $-$0.99999  & $2.55\times 10^{-7}$ & 0.02541 & $1.53\times 10^{-13}$\\
100 &  $-$0.99999  & $2.55\times 10^{-7}$ & 0.02580 & $1.51\times 10^{-13}$\\
\hline
\end{tabular}
\end{table}
According to values in Table~\ref{table:03} the range of eigenvalues is as expected.  The largest eigenvalue is 0.02580 which is of the same order of magnitude, but less than, $c_{mb}$.  The values are very sensitive to the time interval.  Using $\tau_l=0.05$ instead makes the largest eigenvalue in the $n=2$ case negative.    After the eigenvalue $\lambda$ and corresponding eigenvector \textbf{c} have been determined, we calculate the ``error" $e_1=|M\mathbf{c}-\lambda\mathbf{c}|$ ($e_1^{(l)}$ for the lowest eigenvalue and $e_1^{(h)}$ for the highest (most backflow).) It turns out that the error for the largest eigenvalue is much smaller than for the smallest eigenvalue.  It may be that the smallest eigenvalue has other eigenvalues nearby leading to greater difficulty in isolating it.  It is conjectured that in the Bracken and Melloy model there may a discrete spectrum in the interval $(0,c_{mb})$ and a continuous spectrum in the interval $(-1,0)$~(see Ref.~\cite{yearsley12}).  We are not certain whether there is more than one positive eigenvalue.  Overall there are $n$ eigenvalues in the interval $(-1,c_{mb})$ and as $n$ increases the density of eigenvalues increases.  One speculates that the density in the $(-1,0)$ range is higher than in the $(0,c_{mb})$ range.

Of greater importance is the time interval over which the backflow probability is calculated,  since the results are sensitive to the time interval.  For the free particle case the time interval involves small times, for which the nonescape probability is still in the vicinity, but less than, one.  We have seen in the $\delta$-function barrier case that after long times when the nature of decay changes from an exponential to an inverse power law, there is backflow with probabilities less than $10^{-8}$.  Clearly to obtain substantial backflow we need to consider short times.  Whether we can achieve values close $c_{mb}$ is an open question, as is the question whether there is a limit which is equal to $c_{mb}$.  As we increase the value $n$ the backflow probability saturates.  There is a small difference between the backflow for $n=50$ and $n=100$.  

\begin{figure}[h]
\centering
\resizebox{1\linewidth}{!}{\includegraphics[width=.5\textwidth, angle=-90]
{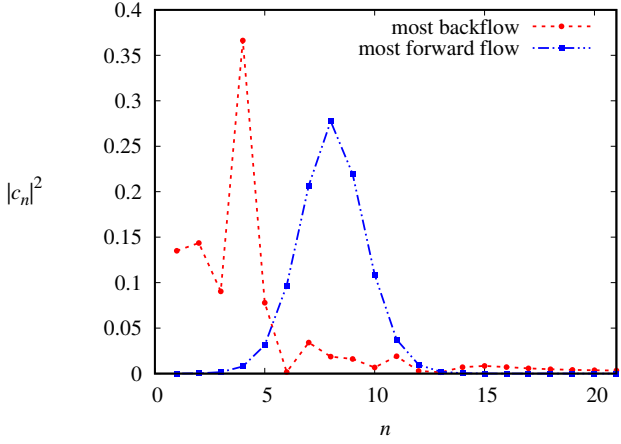}}
\caption{The norm of the coefficients $|c_n|^2$  for the $n=20$ case in Table~\ref{table:03} when backflow is maximum (red round dots), and when forward flow is maximum (blue square dots). \ifdraft{THIS IS DIFFERENT THAN EARLIER GRAPH. PLEASE CHECK.}\fi }
\label{fig:12a} 
\end{figure}

Considering a particular case of $n=20$ we are able to study the composition of the wave function that leads to maximum forward flow or maximum backflow.   In Fig.~\ref{fig:12a} the magnitude of the coefficients $|c_n|^2$ are displayed\ifdraft{\footnote{$\sim$/2019/backflow/calculations/eigenvalue\_v7d.f}}\fi.
The largest backflow occurs with a wave function constituted of components with $n<6$ and a maximum component when $n=4$, whereas the largest forward flow occurs with components with $n<11$ and a maximum at $n=7$.  For both cases components with $n>11$ do not appear to contribute significantly.

\begin{figure}[h]
\centering
\resizebox{1\linewidth}{!}{\includegraphics[width=.5\textwidth, angle=-90]
{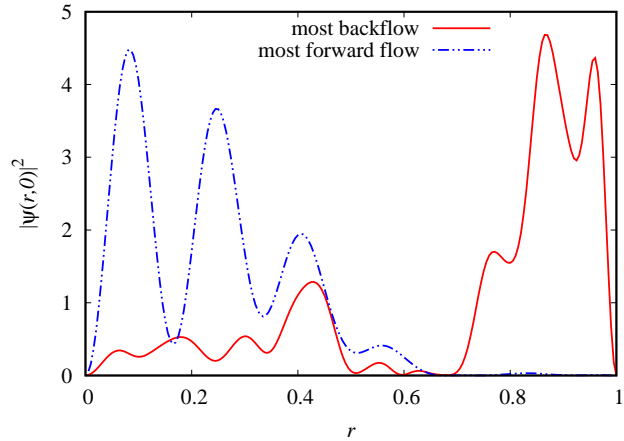}}
\caption{The probability density at $\tau=0$ for the $n=20$ case in Table~\ref{table:03} when backflow is maximum (red solid curve), and when forward flow is maximum (blue dashed curve).}
\label{fig:13a} 
\end{figure}

The initial wave function for the two cases when $n=20$ is shown in Fig.~\ref{fig:13a}.
The initial wave function is localized in the interval $(0,1)$; when maximum backflow occurs it seems to be concentrated close to the endpoint $r=1$, and when maximum forward flow occurs the initial wave function is primarily in the first half of the interval, oscillating with decreasing amplitude. 

\begin{figure}[!h]
\centering
\resizebox{1\linewidth}{!}{\includegraphics[width=.6\textwidth, angle=-90]
{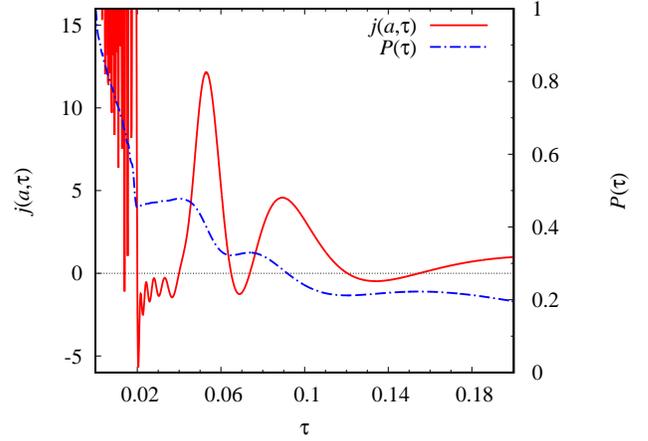}}
\caption{The current probability density at $r=a$ (red solid curve) and the nonescape probability (blue dashed curve) as functions of $\tau$  for the $n=20$ case in Table~\ref{table:03} yielding maximum backflow.}
 \label{fig:14a} 
\end{figure}

In Fig.~\ref{fig:14a} the current probability density at $r=a$ and the nonescape probability are plotted as functions of $\tau$ for the $n=20$ case.  The slope of the nonescape probability curve is positive during the time intervals when the probability current density is negative.  The total time shown spans more than two half-lives, but the decay is clearly nonexponential.  Since we used the parameters of Table~\ref{table:03}. i.e., $\tau_l=0.02$ and $\tau_u=0.04$, we note that the probability current density is prominently negative in that time interval. 
 Exploring that region in $\tau$ space, we find that the interval $(0.020,0.036)$ gives slightly more backflow, i.e., $\lambda_\mathrm{max}=0.02712$, which seems to be the maximum value in that time region.  Taking a larger region e.g.,$(0.02,0.06)$, we obtain $\lambda_\mathrm{max}=0.01757$.  This time interval encompasses the backflow as well as significant positive probability current density, hence the net backflow is smaller.   One expects the greater amounts of backflow to occur shortly after initial decay since the amplitudes of the probability current and density decrease significantly in time.  For example in the time interval $(1.00,1.40)$ we reach a maximum backflow probability of $5.971\times 10^{-8}$. 
 
\begin{figure}[!h]
\centering
\resizebox{1\linewidth}{!}{\includegraphics[width=.6\textwidth, angle=-90]
{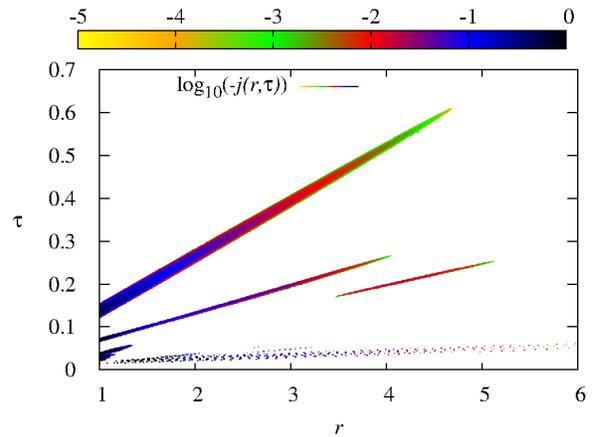}}
\caption{The logarithm of the absolute value of the current probability density where it is negative as a function of $(r,\tau)$  for the $n=20$ case in Table~\ref{table:03} yielding maximum backflow.}
 \label{fig:15a} 
\end{figure}

Figure~\ref{fig:15a} shows the regions of negative probability current density\ifdraft{\footnote{$\sim$/2019/backflow/calculations/eigenvalues\_v7e.f}}\fi    ~in the $r\tau$ plane for the free particle $(n=20)$ wave packet with a maximum backflow over the $0.02$ to $0.04$ time ($\tau$) interval.
The elongated shapes with positive slopes indicate an outward movement of these regions.

\section{Summary and concluding comments}
\label{sec:06}

In the decay of quasistable systems there are time intervals during which the nonescape probability increases and the probability current density is inward.  In the case of the finite $\delta$-shell potential with $\phi_n(r), n=1$, as initial wave function, this backflow occurs after many half-lifes when the decay makes a transition from exponential to inverse-power law behaviour and the nonescape probability is very small.   However, surprisingly the amplitude of the negative probability current density increases as one moves further away from the barrier.  The corresponding backflow amounts are very small however.  These quantities can be calculated accurately because we have access to the exact wave function of this model at any $r$ and any $\tau$.  A regular pattern of backflow results as seen in Fig.~\ref{fig:10}. 

  For the free particle with an initial wave function $\phi_n(r)$ at a single $n$, we do see somewhat erratic behaviour of the probability current density in the transition region from exponential to inverse-power law, but not sufficient to detect backflow.  When the initial wave function is a superposition of $\phi_n(r)$ with different $n$, we obtain substantial backflow at very short times.  Clearly interference of components with different values of $n$ play  a role  as transitions occur between regimes dominated by different $n$ components.  In other words, the shape of the initial wave function can have a profound effect on the amount of backflow.  A similar study with an initial function with different $n$ components for the potential barrier problem may also result in short-time backflow, but that is beyond the scope of this paper and left as future work.

Traditionally quantum backflow involves wave packets whose momentum components are truncated in momentum space, e.g., nonzero components exist only for positive momenta.  The backflow studied in this paper involves wave packets which are initially limited in coordinate space, i.e., only nonzero for $r\in(0,a)$ at $\tau=0$.  Whereas the momentum composition does not change in time, the spatial extent of the wave packet, that is initially localized in space, changes significantly.    
 
Although we show that backflow is present in decaying systems, with or without interactions, it is very small in the decay through a $\delta$ barrier that we study.  For the free particle case we obtain backflow probability of around 2.58\%.  This is less than $c_{mb}=3.8452$\% obtained in the original backflow analyses; however there is no analytical account of this limit~\cite{yearsley12}.  With the model of the free particle of this paper the value more or less saturates and is not expected to increase significantly by including a greater number of states.  The time interval chosen plays an important role in this model, but not to the extent of altering the maximum backflow substantially.  It must be emphasized that backflow occurs for the potential barrier model and for the free particle case, but the interference causing it has different origins, \textit{viz.}, exponential and inverse-power law for the $\delta$ barrier, and two or more different exponential regions for the free particle. 

It would be interesting to investigate the free particle ``decay" using Bohmian mechanics as we did for the $\delta$ barrier~\cite{nogami00,toyama03}.  Initial work on this is promising.

The effect of quantum backflow on transparent boundary conditions is another area that invites scrutiny.  The use of transparent boundary conditions is an approach~\cite{baskakov91} frequently employed in numerical calculations to account for the effect on the propagating wave function due to the boundary of the computational space.  The method referenced~\cite{baskakov91} is claimed to be exact, but yet in its derivation terms are neglected with justification that varies from author to author.  In any case the mechanism by which the wave packet is allowed to cross a boundary without reflection when substantial backflow occurs, is worth investigating. 

\begin{acknowledgments}
The authors thank Professor Y. Nogami for helpful discussions and Professor R.K. Bhaduri for constructive comments on an earlier draft of the paper.
WvD is grateful for the hospitality of Kyoto Sangyo University where part of the research was done.  
\end{acknowledgments}

\ifdraft{\bibliographystyle{apsrev}}\fi    
%

\end{document}